\documentclass[12pt,preprint]{aastex}
\newcommand{\kmps}{km~s$^{-1}$}
\newcommand{\Rsun}{\hbox{$R_{\sun}$}}
\newcommand{\Msun}{\hbox{$M_{\sun}$}}

\newcommand{\K}[1]{\hbox{$K_{\rm #1}$}}
\newcommand{\M}[1]{\hbox{$M_{\rm #1}$}}
\newcommand{\Rhox}{\hbox{$\int$\kern-0.05em$\rho$dx}}
\newcommand{\vsini}{\hbox{$v$$\sin$$i$}}
\newcommand{\ibvs}{IAU Inf.\ Bull.\ Var.\ Stars}
\newcommand{\obs}{Obs.}

\shorttitle{Spurious Eccentricities}
\shortauthors{Eaton}

\begin{document}

\title{SPURIOUS ECCENTRICITIES OF DISTORTED BINARY COMPONENTS
}

\author{Joel A. Eaton}

\affil{Center of Excellence in Information Systems,\\
Tennessee State University,\\
Nashville, TN}

\email{eaton@donne.tsuniv.edu}

\begin{abstract}
I discuss the effect of physical distortion on the velocities of close binary components 
and how we may use the resulting distortion of velocity curves to constrain some properties 
of binary systems, such as inclination and mass ratio.  Precise new velocities for 5~Cet 
convincingly detect these distortions with their theoretically predicted phase dependence.
We can even use such distortions of velocity curves to test Lucy's theory of convective 
gravity darkening.  The observed distortions for TT~Hya and 5~Cet require the contact 
components of those systems to be gravity darkened, probably somewhat more than predicted 
by Lucy's theory but clearly not as much as expected for a radiative star.  These results 
imply there is no credible evidence for eccentric orbits in binaries with contact components.  
I also present some speculative analyses of the observed properties of a binary encased 
in a non-rotating common envelope, if such an object could actually exist, and discuss 
how the limb darkening of some recently calculated model atmospheres for giant stars may bias
my resuts for velocity-curve distortions, as well as other results from a wide range of 
analyses of binary stars.  
\end{abstract}


\section{INTRODUCTION}

Two recent papers suggest the time is ripe to look critically at the effect 
of stellar distortion on velocities of variable stars.  In the first, Miller 
et al.\ (2007) found a slightly eccentric orbit for the Algol binary TT~Hya, 
a system that has gone through mass exchange and, in so doing, surely 
circularized its orbit.  In the second, Wood, Olivier, \& Kawaler (2004) used 
velocity variations calculated for a rotating prolate spheroid as a possible 
mechanism for explaining long-period velocity variations of AGB stars.  The 
first paper assumes these distortions are irrelevant; the second, that they 
are crucial.  

Solutions of the velocity curves of close binary systems often have small 
eccentricities thought to be spurious.  Lucy \& Sweeney (1971, 1973) argued 
that most of these were statistical flukes resulting from errors of measurement.  
However, as we shall see in \S~3, there are circumstances in which such 
eccentricities result from interesting physical processes, and for which the 
resulting distortions may be used to extract useful information about the mass 
ratios and inclinations of these systems.  Such distortions from proximity 
effects were actually predicted by Sterne (1941), explored by Wilson \& Sofia 
(1976) as a way of determining mass ratios and inclinations of X-ray binaries, 
and discussed in passing by Wilson (1979), but otherwise they do seem to 
have been ignored as a way of extracting information about binaries.

Red giants often have unexplained secondary periods much longer than their 
radial fundamentals (e.g.\ Wood et al.\ 2004).  They manifest uncomfortably 
large radial-velocity variations on those periods (Wood et al.; Hinkle et 
al.\ 2002), which imply pulsational excursions of the order of 30\% of the 
radius of the star.  These variations remain one of the burning mysteries 
of stellar astronomy (Derekas et al.\ 2007).  They seem to have the sort 
of color variations expected of radial pulsation, yet there is no known 
mechanism for giving such long periods.  

I shall explore the implications of distortion on observed radial velocities, 
starting with an assessment of the rotating spheroids of Wood et al.\ in \S~2.1, 
then exploring how a binary encased in a common envelope might appear to us 
in \S~2.2, and finally discussing the much less speculative effects of distortion 
on radial velocities of close binary components in \S~3.

\section{ROTATION OF A PROLATE SPHEROID}

Wood et al.\ (2004) used a prolate spheriod rotating about its short axis 
to model velocity variations of AGB stars, although they ultimately rejected 
that idea.  Since such a figure will project symmetrically onto the sky about 
its rotation axis, there would be no net velocity variation for a uniformly 
illuminated disc.  So these variations must occur solely from the effects 
of limb and gravity darkening over the distorted surface (P.\ Wood, private 
comm.\ in 2007; Sterne 1941). 

Let us make some standard assumptions about the rotating spheriod.  Give it axes 
{\it a} and {\it b} in its equatorial plane and, for simplicity, {\it b} toward 
its rotational pole.  Assume the mass is concentrated to its center, so the local 
gravity is proportional to 1/$r^2$, and let the gravity darkening be convective, 
F$_{\rm Bol}$ $\sim$ $\nabla\Omega^{g}$, with $g$=0.32 following Lucy (1967), 
where $\Omega$ is the gravitational potential.  Assume total darkening to the limb 
($x$=1.0), a safe assumption for such cool stars.  Finally, assume that the radial 
velocity of the centroid of light is the radial velocity measured for the star.  
Now, if we write a computer program for these somewhat questionable assumptions, 
we can calculate the light and velocity curves for such a star, if any such object 
were to exist.

Figure 1 shows some velocity curves for a prolate spheriod with $a$=0.50 and 
$b$=0.25, a rather extreme case that illustrates the model pretty well.  For 
this combination of parameters, we get light variation of about 1.9 mag at 
$i$=90$\degr$, somewhat larger than seen in the AGB stars.  The solid and dashed curves 
have $x$=1.0 with $g$=0.0 and $x$=0.0 with $g$=0.32, respectively, to show the separate 
effects of limb and gravity darkening, which seem to cancel each other out in this 
calculation.  The dotted curve, for $x$=1.0 with $g$=0.32, shows the combined effect.  
Both limb and gravity darkening obviously cause significant velocity variations mimicing 
an eccentric orbit.  The curve for pure limb darkening in Figure 1 may be fit with 
an orbit in which $e$=0.21 and $\omega$=270$\degr$.  The case of pure gravity darkening 
gives $e$=0.22 and $\omega$=87$\degr$.  Other aspect ratios ($b$/$a$) give similar results 
but smaller amplitudes and somewhat smaller eccentricities for less elongation.  More 
elongation gives very slightly bigger eccentricities. 

These calculations for a rotating prolate spheroid do not represent the sort of 
long-period velocity variation observed in AGB stars.  The eccentricities we find for 
this ellipsoidal model $e$ $\sim$ 0.2) are significantly smaller than the the formal 
eccentricities of cool giants when compared to either the largest values measured or 
typical ones.  Fitting the velocity shifts observed for 9 stars with elliptical orbits 
gave $e$ in the range 0.08--0.49 and $\omega$ in the range 195--418\degr\ 
(Hinkle et al.\ 2002; Wood et al.\ 2004), with medians $e$=0.33, $\omega$=251\degr, 
and $K$=2.3 \kmps.  The phase dependence of the calculated theoretical effect for 
gravity darkening is wrong in that it gives a rapid drop in velocity, as in a classical 
radial pulsation, not in the observed rapid rise.  Furthermore,  the velocity variation 
in these calculations ($\Delta$RV/\vsini\ $\sim$ 0.04) is small enough to require a 
rotational velocity (58 \kmps\ for $K$=2.3 \kmps) prohibitively large for AGB stars, 
for which $v_{\rm rot}$ $\lesssim$ 3~\kmps\ (Olivier \& Wood 2003).

\subsection{Speculations about an Encased Binary}

It is not clear that any star, especially an AGB star, would be a prolate spheroid 
rotating about its fixed center of mass.  The only way I can see to get a significant 
elongation is through tidal distortion in a close binary system, and the effects of 
that distortion on the velocities can be rather subtle, as we shall see in \S~3.
Wood et al.\ argue that some of these AGB stars may be coalesced binaries, presumably 
with double cores encased in a common envelope to account for the prolate shape, a 
radically imaginative idea deserving a closer look.

The size of such an object presents another fundamental problem with trying to use a 
rotating prolate spheroid to explain the light and radial-velocity variations of AGB 
stars.  If the star has a double core, its period and linear scale are fixed by 
Kepler's laws.  Consider, for example, a star with a double core having a period of 
1588 d (4.35 yr) as discussed by Wood et al.\ for Z~Eri.  It would 
have a semi-major axis of 3 au (640 \Rsun) for two 0.75~\Msun\ components.  Even in 
a binary consisting of two 1.0~\Msun\ components with a period of 750~d, the semi-major 
axis would be about 440\Rsun.  The radii of these objects would be of the order of 
their semi-major axes.  Wood et al.\ estimate radii near 170 \Rsun\ for the stars 
with long secondary periods, so encased binaries with the right periods to explain 
the observed long-term velocity variations would seem to be too large.  Also, if 
these stars were synchronously rotating, admittedly a completely unphysical condition, 
our two examples would have $v_{\rm rot}$ of 20 and 30 \kmps, respectively, much larger 
than the \vsini's of AGB stars.

The idea of a binary totally encased in a common envelope seems preposterous, since 
the gravitational equipotential surfaces above the two cores, which presumably help 
control the density structure of the star, become very complicated with three 
Lagrangian points at which the gravitational acceleration vanishes (e.g., Kopal 1959, 
Figs.\ 7-7 through 7-10).  However, this may be no worse than the generally accepted 
idea of a contact binary containing a single such Lagrangian point.  Given these 
circumstances, however, it is not clear to me that any star could be stable with a 
surface bigger than in a typical contact binary (one with its surface between the 
first and second Lagrangian surfaces).  

A more fundamental problem with this idea is the question of solid-body rotation.  
For a synchronous binary bigger than its second Lagrangian surface, the centrifugal 
potential dominates, and the surfaces of constant potential (normally the level surfaces) 
are unbound.  Any star containing a double core must, therefore, be rotating {\it slower 
than synchronously} in its outer layers.  I shall leave it to others to estimate the time 
the churning going on inside such an object would take to dissipate the angular momentum 
of the encased core.  However, in this case, second-order effects of rotation cannot dominate 
the light or radial velocity of the star.  Instead it must be subject to a non-radial 
pulsation driven by the encased binary.  In this regard it is no different than other 
non-synchronous binaries, such as the components of eccentric systems discussed by 
Wilson (1979).

Now, in the remote possibility that binaries with an encased core can actually exist 
long enough for us to observe them, just what might they look like?  We can make some 
simplifying assumptions that let us calculate light and velocity curves for such an 
object.  As a first stab, I have considered a system that is not rotating at all.  This 
eliminates the centrifugal potential and makes the gravitional potential simple:
\begin{equation}
     \Omega = 1/r_1 + q/r_2,                                  \eqnum{1} \\
\end{equation}
where $r_{\rm 1}$ and $r_{\rm 2}$ are the distances of some point from the two components 
of the binary, the origin of coordinates is the binary's center of mass, and $q$=\M2/\M1 
is the mass ratio of the system.  The surface corresponding to an equipotential of this 
equation is shaped roughly like a prolate spheroid for equal masses ($q$=1.0) and a radius
large enough.  For other mass ratios, the shape is more distorted and becomes rather 
biological at smaller radii.  We can calculate the light and velocity curves for such 
objects if we make an assumption about how surface brightness changes over the star.  I 
have made the simplifying assumption that the surface is always given by an equipotential, 
that this figure rotates with the binary system (i.e., has no phase lag), and that the 
surface brightness is determined by limb and gravity darkening as in a normal synchronously 
rotating binary.  Actually, the surface brightness would be determined by the driven 
pulsation, so these assumptions are clearly bogus.  However, they do give us a way to 
estimate the effect of having an encased binary as a star's core, at least to first order.  

For these assumptions we may calculate the pulsational velocities, assumed radial, by 
comparing points equal distances apart in azimuth and dividing by the time it takes the 
binary to rotate that amount.  Since in the calculations, we divide the star's surface 
into a number of sectors of azimuth, each $\Delta\lambda$ wide in radians, the local 
pulsational velocity along the star's radius is 
\begin{equation}
     V_{j} = { { {d\lambda}\over{dt} } { {dr\over{d\lambda} } = { {2\pi}\over{P}} { {a(r_{j+1} - r_{j-1})}\over{2\Delta\lambda} } }i },       \hfill                       \eqnum{2} \\
\end{equation}
where $a$ and $P$ are the semi-major axis and period of the encased binary, $r_j$ is the 
radius in units of $a$, and index $j$ represents the steps in azimuth.  The spheriodal 
pattern, rotating through the surface of the star, breaks the surface up into four sectors 
with alternating positive and negative pulsational velocities.  These, in turn, project 
into our line of sight in a way that gives the star a roughly sinusoidal velocity variation 
on half the period of the encased binary.

Given the foregoing assumptions, we may calculate the properties of the encased binary to 
first order.  Figure 2 shows roughly what such a star with a mass ratio of 0.33 and a 
radius of 1.3 times the semi-major axis might look like.  Figure 3 shows some light curves 
for this sort of object, as well as for the two limiting contact binaries.  Figure 4 shows 
some velocity curves.  The important thing to note is that, for equal masses, the object gives 
a light variation on half the orbital period of the encased binary.  The radial-velocity 
variation is also on this period.  For different mass ratios, the degeneracy in the light 
variation goes away, but the radial velocities still vary on half the orbital period in a 
peculiar way.  If we ever were to find a star in the common-envelope stage, it would most 
likely have the light variations of a contact binary (e.g., dotted and dot-dashed curves 
in Fig.\ 3) but with velocity variations on {\it half} the photometric period.

\section{SPURIOUS ECCENTRICITIES FROM TIDAL DISTORTION}

The contact components of Algol binaries really are like prolate spheriods (actually, 
triaxial ellipsoids), but only to first order (e.g., Russell 1912).  Instead, they 
have a complicated asymmetric figure if they actually follow the contours of their 
Roche lobes.  These lobes are symmetric about a line between the two component stars, 
so the distortions would be symmetric about conjunctions.  However, they would be 
different for the two conjunctions in a way that distorts the velocity curve.  We 
can predict this distortion with binary modelling software and test it with precisely 
measured velocities of contact components of close binaries (Wilson \& Sofia 1976).

The new date we are using here are radial velocities from the Tennessee State University 
(TSU) Automatic Spectroscopic Telescope (AST; Eaton \& Williamson 2007).  These echelle 
spectra of roughly 30,000 resolution cover the wavelength range 6000--7100 \AA.  
We reduced and analyzed them with standard pipeline techniques to derive the radial 
velocities for 5 Cet (HD~352) and AX~Mon (HD45910) given electronically as Table 1.
Velocities measured for bright stars such as 5~Cet have an external precision of 
$\sim$0.11 \kmps\ RMS (Eaton \& Williamson 2007).  Listed in the table (see printed 
excerpt) are (1) HJD, the Heliocentric Julian Date of observation (minus 2,400,000), 
(2) $RV$, the radial velocity of the cool star, and (3) a tag identifying the star 
by its HD number.

\subsection{Modelling the Velocity Distortions}

We can calculate the effect of tidal distortion combined with limb and gravity 
darkening on the velocities of Algol components if we make an assumption about 
how the measured velocity is related to surface brightness on the star.  I shall 
make the simplifying assumption\footnote{One should keep in mind that there 
may be small, third-order systematic errors introduced by this approach (Winn et al.\ 
2005).} that a line profile is constant over the disc, and that the observed velocity 
is just the local velocity weighted by intensity and averaged over the visible disc.  
For a sphere, the velocity calculated in this scheme is just the velocity at disc center, 
so there are no deviations from the 
velocity of the center of mass.  For a distorted binary component, we can calculate the 
deviation by averaging $y$*$\Delta$$A$*$I_\lambda$ over the component, where $y$ is the 
distance perpendicular to the rotation axis projected into the sky, $\Delta$$A$ is the 
projected area of a surface element in the calculation, and $I_\lambda$ is the intensity 
as determined by limb and gravity darkening.  We can get this information from computer 
programs for calculating light and velocity curves of binaries, such as Wilson's (1979), 
and I have coded it into my own program (e.g., Eaton et al.\ 1993), which is easier for 
me to understand.

Figure 5 shows the distribution of surface elements in a typical calculation 
for phase 0.25 when the bigger, distorted star is approaching.  The second component 
(generally hotter and smaller) is to the right (at $y$=1.0) in this coordinate system.  
At the phase illustrated, the left side of the star (negative $y$'s) contributes more light 
than the right side, in spite of there being more area to the right, because of limb and 
gravity darkening.  Some results are given in Figure 6  where we see the effect 
of the distortion on a typical velocity curve at the top and some calculated deviations 
in the lower panel.  I have expressed these curves in terms of the total velocity 
amplitude of the system (\K1+\K2), since the calculated quantity 
($\sum$$y$*$\Delta$$A$*$I_\lambda$) is the displacement of the centroid of light from the 
rotation axis in units of the separation of the two stars.  Three things are clear in 
Figures 6 and 5.  First, the effect on the velocity curve is subtle but systematic.  Second, 
the magnitude of the effect would depend upon mass ratio mostly through changes in the 
relative velocity amplitudes of the two stars, the less massive the undistorted companion, 
the smaller the effect on the distorted star's velocity curve.  Third, the effect is 
smaller at lower inclination because the roughly symmetric gravity-brightened pole 
contributes more of the light.

Figure 6 shows how the mass ratio affects distortion of the velocity curve.  Figures 
7 and 8 show the effects of inclination and gravity darkening.  Both are significant.  
By contrast, the effect of limb darkening is more moderate\footnote{If we adopt linear 
limb darkening, as I have, a change of \hbox{$\Delta$$x$$_V$}=0.1\ corresponds to 
\hbox{$\Delta$$g$}$\sim$\hbox{$-$0.012} or \hbox{$\Delta$$i$}$\sim$0.8. The effect 
on mass ratio would be rather large, however (see Fig.\ 6).}.  The large dependence 
on inclination in Figure 7 means we can use these velocity-curve distortions to estimate 
inclinations of non-eclipsing Algol binaries.  I have plotted these curves for binaries 
against photometric phase, for which the zero point is superior conjunction of the 
distorted star, the phase of minimum light in ellipsoidal variation (e.g., Hall 1990).

\subsection{Tests for Some Actual Stars}

HD~352 = 5~Cet (also, perversely, AP~Psc) is a close semi-detached binary system 
with a giant contact component (e.g., Eaton \& Barden 1988).  It is rather bright 
and single-lined.  We have obtained 126 observations of it over a period of three 
years with the AST.  The velocity curve deviates from the sinusoidal shape expected 
for a spherical star in a circular orbit by a noticeable amount (see Fig.\ 9).  We 
can reduce the obvious systematic errors of the fit by allowing an eccentricity, 
$e$=0.033$\pm$0.002 with $\omega$=88.3$\pm$0.1.  However, there are still systematic 
deviations from the fitted curve.  Figure 10 shows the deviations from a sinusoid.
They are obviously of the same shape, and if we apply the right values for the 
physical parameters of the system (see \S~3.3), of roughly the right amplitude 
as the theoretical calculations in Figure 6.  You may note that the data scatter 
by roughly the expected external error (0.1 \kmps) about this relation.

The spurious eccentricity we have found for 5~Cet has essentially the same magnitude 
and orbital orientation as Miller et al.\ (2007) claimed for TT~Hya.  Furthermore, the 
velocity curves of the cool components of all Algol binaries should show spurious 
eccentricities of this type ($e$$\sim$0.02--0.04, $\omega$ $\sim$ 90$\degr$).  That they 
do not simply means that the measured velocities have not been precise enough to detect 
this effect, as Lucy \& Sweeney told us.  However, we may test this idea further with 
177 radial velocities from the AST for AX~Mon (HD~45910) for which we have a preliminary 
solution.  In that case both circular and elliptical orbits fit the data to $\sigma$ 
$\approx$ 0.9 \kmps.  The deviations from the circular orbit (Fig.\ 11) have roughly 
the phase dependence of Figure 6 with an amplitude $\sim$4~\kmps\ peak-to-peak.

Now, we might ask, did Miller et al.\ really have data for TT~Hya precise enough to 
detect the spurious eccentricity predicted for rotational-tidal distortion?  I have 
tried fitting the velocities in their Table 2 and get comparable fits for circular 
and elliptical orbits ($\sigma$ $\approx$ 6 \kmps).  If we restrict ourselves to 
their more precise data, those for 1994--2001, we can recover the elements listed in 
their Table 4 (col.\ 2) to within the probable errors.  However, this is {\it only if 
we apply an unacknowledged 9.9 \kmps\ zero-point ``correction" to Peters' data}.  For 
a circular orbit fitted to this ``corrected" data set, the deviations have roughly the 
phase dependence of Figure 6 but with the 2 \kmps\ scatter superimposed.  The amplitude 
expected (for \K1+\K2=167 \kmps, $i$$\sim$83, $q$=4.38, $x$=0.77, and $g$=0.32 per 
Miller et al. [2007] and Van Hamme \& Wilson [1993]) is $\sim$ 8.2 \kmps\ peak-to-peak, 
a bit smaller than the noisy observed value ($\sim$11 \kmps).

\subsection{Some Further Applications}

We may use the calculations in \S~3.1 to restrict some of the parameters 
of Algol systems, given precise data.  The amplitude of the deviations of the 
velocity curve from a sinusoid depends primarily on the orbital inclination, $i$,
the velocity amplitude of the companion star, \K2, and the amount of gravity 
darkening, $g$.  It also depends on limb darkening, but effectively much less than 
on $i$ or $g$, if we can assume the limb-darkening law is known as precisely as usually 
assumed in light-curve analyses
Figure 6 shows that the deviations centered on superior conjunction 
(phases 0.75--0.25) change much more with inclination than those centered on inferior 
conjunction (phases 0.25--0.75).  That dependence provides a way of restricting 
inclinations of the non-eclipsing systems, such as 5~Cet and AX~Mon, if we know $g$.  
In the eclipsing systems, on the other hand, we would get a moderately weak constraint 
on the mass ratio through \K2, although in these systems $q$ is usually very well 
constrained by the light curve.  Wilson (1979, \S~VI) would argue that these constraints 
are built implicitly into his light-curve solutions.  However, by stripping them out, 
we can isolate them and look at just what they might tell us about various individual 
properties of a star and the physical processes that apply.

5~Cet gives us a good example of applying these velocity-curve distortions.  In this 
system (P=96.4 d, \K1=24 \kmps, and $q$$\leq$0.82, whence \M1sin$^3i$ $\geq$ 1.50 \Msun) 
the inclination is probably fairly small to get reasonable masses.  We previously estimated 
$i$$\sim$60\degr\ (Eaton \& Barden 1988).  For these present best guesses, the distortions 
predicted are somewhat smaller than observed (solid curve in Fig.\ 10).  For AX~Mon, 
we may get estimates of the properties of the system from Elias et al.\ (1997), 
namely \K1+\K2=66--70 \kmps, $i$=50--70\degr, and $q$=4.0--3.0.  For values characterizing 
this range ($i$=60\degr and $q$=3.0, whence \K1+\K2=69 \kmps), the calculated curve for 
$g$=0.32 is much too shallow to fit the observations.  Even if we increase $g$ to 0.5, the 
inclination is too small.  Instead, the inclination is probably larger, or the gravity 
darkening even larger than it seems for 5~Cet.  Thus the velocity-curve distortions 
constrain the properties of this non-eclipsing binary to the ends of the ranges expected 
from photometry, namely to $q$$\approx$3.0 and $i$$\approx$75\degr.

The rather strong differences of the curves in Figure 8 gives us what is perhaps a 
unique test of Lucy's (1967) theory of gravity darkening for convective stars.  
This brilliantly simple theory predicts $g$=0.32 for convective stars; more detailed 
calculations by Claret (2000, 2003) for a wide range of convective stars predict 
essentially the same level of gravity darkening or slightly less.  A simple comparison 
of the observed velocity-curve distortions in Figures 10--12 with the theoretical curves 
in Figure 8 shows the binary components must be gravity darkened.  On the other hand, 
the relative amplitudes of the effect for the two conjunctions in the observations is 
much lower than for {\it radiative} gravity darkening ($g$=1.0).  The amount of gravity 
darkening is hard to measure because of the residual scatter in the data and because 
two of our three stars have indeterminate inclinations.  However, in Figure 10 the shape 
of the observed curve is much closer to the shape calculated for $g$=0.52 than for 0.32.  
In Figure 11, as we have seen, a similar increase in $g$ gives the observed amplitude 
without requiring $i$ to be unacceptably large.  In Figure 12, for a system in which 
the geometry is very highly constrained, $g$=0.52 fits the amplitude much better than 
$g$=0.32 and $g$=1.0 does not fit it at all.  So, the amount of gravity darkening seems 
to be around $g$=0.5, a little larger than Lucy's $g$=0.32.  It should be possible to 
constrain the gravity darkening of convective stars much better in future with precise 
radial velocities for more eclipsing stars. 

Tidal distortion should give a variable \vsini\ for a binary component because 
the disc is actually broader at some phases than at others.  We saw this effect 
in UU~Cnc (Eaton, Hall, \& Honeycutt 1991).  The new data for 5~Cet provide 
a further test, and I have applied it by looking at composite spectra of 5~Cet 
at the two conjunctions and at the two quadratures.  First, blends of Fe lines near 
6575 \AA\ and 6594 \AA, which can give a very good measure of \vsini\ without 
numerically measuring and calibrating line widths (Eaton 1990), are somewhat sharper 
at the conjunctions than at the two quadratures.  Second, widths of Gaussians fit 
to cross-correlation functions (of a list of strong solar absorption lines with the 
observed spectrum -- Eaton \& Williamson 2007) for these four composite spectra are 
about 5\% broader for the quadratures than the conjunctions.  See Table 2 for the 
details.  We may simulate this effect by calculating synthetic spectra and measuring 
the variation of line width with phase, and I have done so for a rough model for the 
system.  This model assumes the visible giant star is a contact component with 
$x$=0.70 and $g$=0.32, and that \K1=23.9 \kmps\ and $q$=0.8.  I used a spectrum 
of $\delta$~Eri (K0~IV) from the National Solar Observatory covering the wavelength 
range 6400--6480 \AA\ to represent the cool component in this calculation, but the 
resolution is similar to that in the AST observations.  The calculated line widths 
(col.\ 3 of Table 2) have roughly the same phase dependence as shown by the 
observations but are about 10\% narrower.  We can make them as large as the observed 
widths by reducing the mass ratio to about $q$=0.7, thereby raising \K1+\K2.  In any 
case, a mass ratio greater than 1.0 (unseen star more massive than the contact component) 
gives profiles much narrower than observed.

\section{SUMMARY AND DISCUSSION}

The existence of precise radial velocities for long-period binaries, made possible by 
a dedicated instrument on a robotic telescope, has made it possible to demonstrate the 
distortion of the radial-velocity curves of close binaries predicted by Sterne (1941).
The distortions depend on a binary's inclination and mass ratio in ways that let us 
estimate these properties in favorable circumstances.  Given the nature of this 
velocity-curve distortion, we must conclude that there is no credible evidence for 
orbital eccentricity in binaries with contact components.  Furthermore, the distortion 
observed in three systems requires the stars to be gravity darkened, probably somewhat 
more than expected from Lucy's (1967) theory but much less than expected for stars with 
radiative envelopes.  

In addition to the foregoing substantive results, I have also (1) detected the 
variation of \vsini\ caused by tidal distortion in 5~Cet and used it to derive 
an improved mass ratio, $M_{\rm unseen}$/$M_{\rm gK}$ $\approx$ 0.70, (2) estimated 
somewhat improved geometrical properties for AX~Mon, and (3) explored what properties
a binary encased in a common envelope would have if we ever were to detect one. 

\subsection{A Caveat about Limb Darkening}

The greatest uncertainty in the results of this paper, as well in many others 
of the same genre, comes from the effect of limb darkening.  This uncertainty 
would impact both papers analyzing light curves and those based on rotational 
line broadening or Doppler images.  Such analyses usually parameterize the 
emergent intensity as a function of the angle between the surface normal and 
the line of sight, more specifically on $\mu$, its cosine.  The simpliest
formulation, a linear dependence on $\mu$, follows from the stratification of 
a simple hydrostatic atmosphere in radiative equilibrium.  Other more complicated 
formulations are possible with increasingly less intuitive mapping into the 
atmospheric structure.  Eventually, these effectively seem to become mere fitting 
schemes (e.g., Brown et al.\ 2001).  Alternatively, one may apply a calculated 
intensity directly, perhaps by looking it up in a table of $I_\lambda$ vs. $\mu$.

Theoretical model atmospheres predict levels of limb darkening that may be 
applied in analyses of binary stars and other problems.  Such analyses usually 
assume limb-darkening coefficients from a few standard theoretical lists (e.g., 
Al-Naimy 1978; Claret \& Gimanez 1990).  Normally, such models assume 
plane--parallel stratification in hydrostatic equilibrium, basically the physics 
we apply to the Sun.  A good example is the grid of atmospheres Kurucz calculated 
with his program ATLAS (e.g., Kurucz 1995).  A recent alternative set of models 
is NextGen (Hauschildt, Allard, \& Baron 1999; Hauschildt et al.\ 1999).  This 
latter grid ought to give a superior interpretation of our standard physical 
assumption--for example, by calculating line blanketing in the blue--UV through 
opacity sampling.  Limb darkening from NextGen, especially from the spherical 
models for giants and supergiants (Claret \& Hauschildt 2003), is quite different 
than from plane-parallel models.  It tends to be much more pronounced, with the 
intensity dropping to zero well before the limb (see Orosz \& Hauschildt 2000, 
Fig.\ 4).

If real stars really have the limb darkening of these spherical NextGen models, 
there would be systematic errors in a wide range of observational results for 
giants and supergiants.  For instance, it would make direct measurements of 
stellar diameters systematically small, leading to effective temperatures too 
high.  This biases the Barnes-Evans relation (Barnes \& Evans 1976) used 
extensively to derive angular diameters for cool stars and otherwise calibrate 
the flux--temperature relationship.  This effect might be tested by analyzing the 
consistency of temperatures derived by fitting spectral energy distributions (e.g., 
Bertone et al.\ 2004) with those derived from measured bolometric fluxes and angular 
diameters.  It would likewise bias rotational velocities from line broadening and a 
wide range of properties of binaries derived from such broadening such as mass ratios 
of non-eclipsing systems, as discussed by Orosz \& Hauschildt (2000) and Shahbaz 
(2003).  It would also bias the results for {\it eclipsing} systems by making it harder
to fit the eclipse shapes and ellipsoidal variation simultaneously.  Furthermore, it 
might well bias the Doppler images of spoted stars, if the effects are as extreme as 
calculations of Claret \& Hauschildt (2003, see Fig.\ 2) imply.  In our present 
analysis, the effect of the nontraditional limb darkening of NextGen models would 
likely be to increase the amplitude of the velocity variation somewhat, allowing for 
a lower gravity darkening.  We can see this effect by making the linear limb-darkening 
coefficient $x$$>$1.0 and truncating the intensity at zero when it becomes negative 
near the limb.  Such a calculation does increase the deviation from a sinusoidal 
velocity curve.

Now then, how much do we have to worry about the existing body of lore for stellar 
astronomy?  Perhaps not as much as one would fear.  There are few highly precise 
determinations of limb darkening for cool stars other than the Sun, and that is 
especially true for giants and supergiants.  However, one reliable determination
for a K giant (Fields et al.\ 2003) finds limb darkening much less extreme than
predicted by NextGen models.  Predictions of ATLAS were somewhat closer, but did 
not fit as well as one might hope.  Fields et al.\ suspected unmodelled physical 
effects are to blame for this discrepancy, inasmuch as there are obvious effects in 
atmospheres beyond our current understanding.   None of these models, for instance, 
incorporate the inhomogeneities that we expect in all stellar atmospheres on the 
basis of solar granulation and Ayres's (e.g., 2002) work on CO.  Furthermore, all 
these calculations necessarily use very simplistic models for turbulence, whatever 
that really is, and for convection.  At this point, it seems the theoretical limb 
darkening of NextGen models is not tested well enough for us to worry seriously 
about its effect on binary-star analyses, but that we as a community should take 
it seriously enough to look for more tests, especially for the supergiants.

\acknowledgments 

I would like to thank Peter Wood for stimulating me to think about this problem, for 
giving me a key reference to work on the long-period variations of AGB stars, and
for clarifying some details of his work on rotating spheriods.  This research used 
the SIMBAD database and was supported by NASA grants NCCW-0085 and NCC5-511 
and by NSF grants HRD~9550561 and HRD~9706268.

{\it Facilities:} \facility{TSU:AST}\

\clearpage

%%Table 1 -- Sample Spectroscopic Data

\begin{deluxetable}{rcc}
\tablecaption{Spectroscopic Data}
\tablewidth{0pt}
\tablehead{
\colhead{HJD}          & \colhead{$RV$} & \colhead{Star} \\
\colhead{(2,400,000+)} & \colhead{(\kmps)}         & \colhead{}     
}
\startdata
 52897.9971&  -8.25& HD~352 \\
 52900.9680&  -3.95& HD~352 \\
 52923.9281&  22.48& HD~352 \\
 52959.6797& -14.17& HD~352 \\
 52970.7973& -23.54& HD~352 \\
 52971.7946& -23.58& HD~352 \\
\enddata
\tablecomments{This excerpt shows the form and content of the full table
as published in the electronic edition of the Astrophysical Journal.}
\end{deluxetable}

%%Table 2 -- Line Widths for 5 Cet

\begin{deluxetable}{lcc}
\tablecaption{Some Measured Line Widths for 5~Cet}
\tablewidth{0pt}
\tablehead{
\colhead{Orbital} & \colhead{$\sigma_{\rm meas}$} & \colhead{$\sigma_{\rm calc}$} \\
\colhead{Phase}   & \colhead{(\kmps)}             & \colhead{(\kmps)} 
}
\startdata
Quadrature (approaching):             & 21.73&  19.72 \\
Quadrature (receeding):               & 21.71&  19.87 \\
Superior Conjunction (star behind):   & 20.33&  18.98 \\
Inferior Conjunction (star in front): & 20.03&  18.51 \\
\enddata
\tablecomments{The $\sigma$s are the width parameter from a Gaussian fit to the line profile.  
The first value given for each phase comes from profiles in composite spectra of 5~Cet; 
the second value, from theoretically calculated profiles.}
\end{deluxetable}

\clearpage

%%Figure 1 -- Prolate Spheroid

\begin{figure}
\begin{center}
\epsscale{0.80}
\plotone{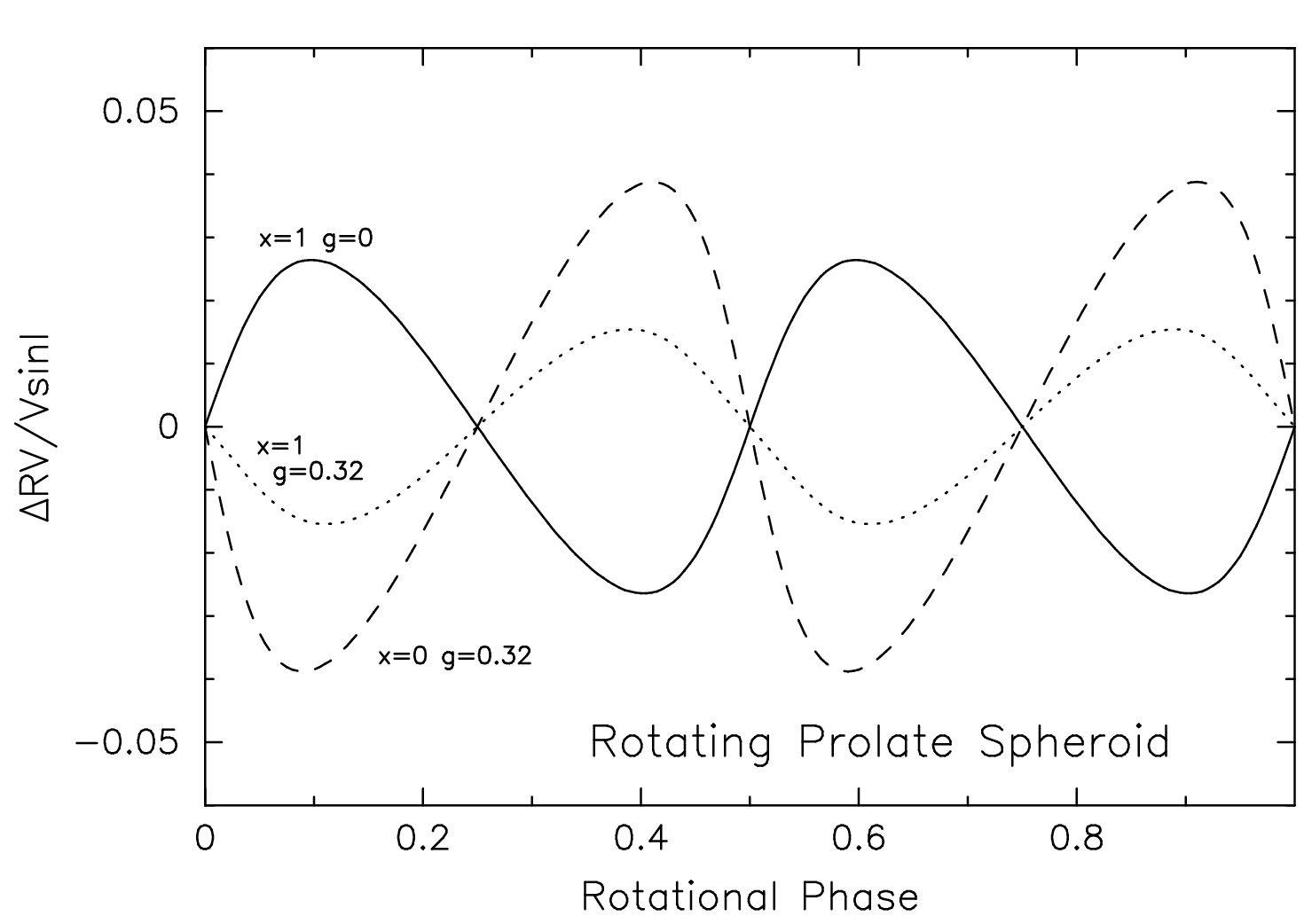}
\end{center}
\caption{Velocity curves of a prolate spheroid with a 2:1 aspect ratio.  The three
curves show the effects of limb darkening alone (solid), of gravity darkening alone 
(dashed), and the combination of limb and gravity darkening (dotted).  All assume 
$i$=90\degr.
\label{fig1}}
\end{figure}

%%Figure 2 -- Surface Plot for Encased Binary

\begin{figure}
\begin{center}
\epsscale{0.80}
\plotone{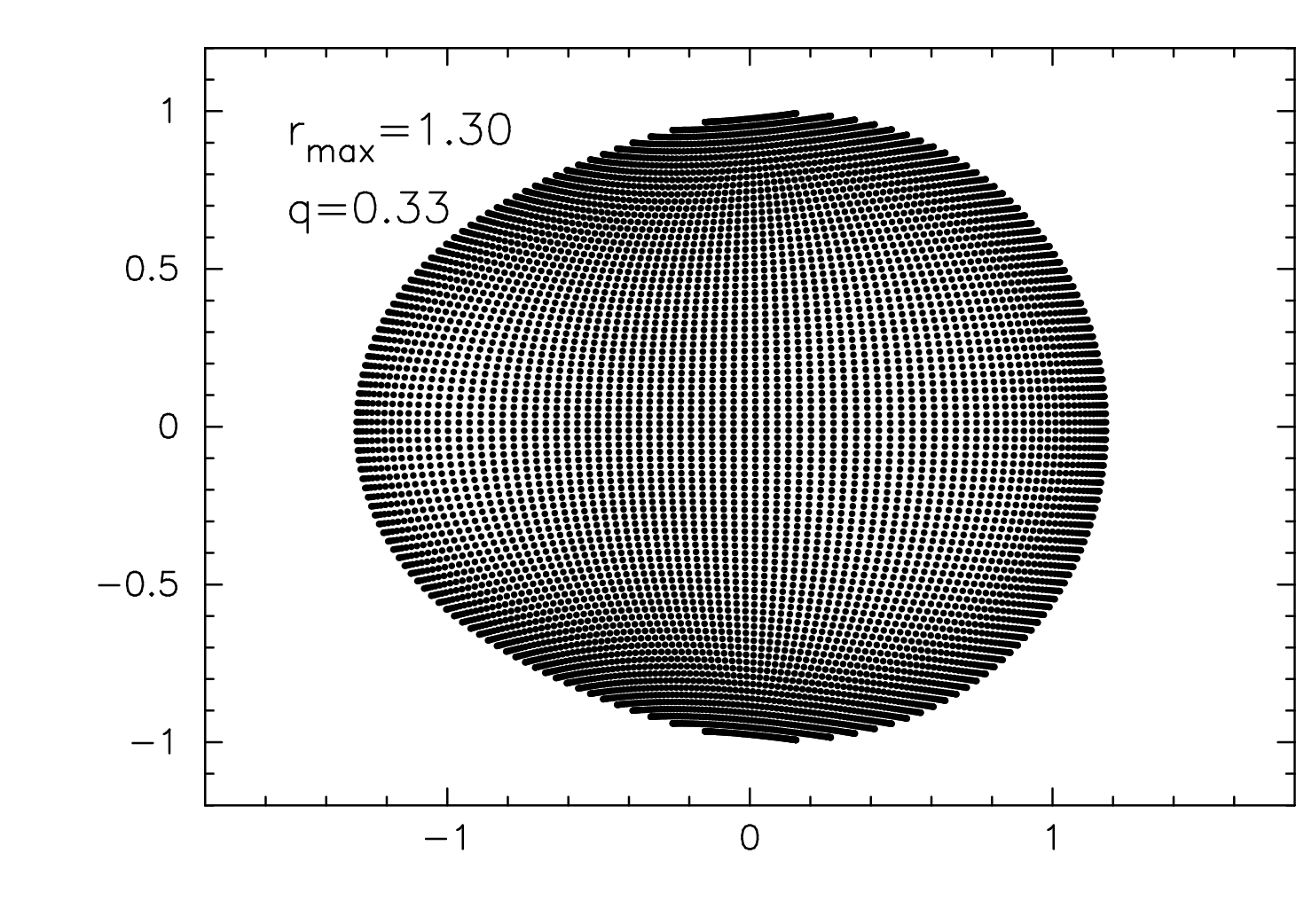}
\end{center}
\caption{The distribution of surface elements in a representative calculation for 
an encased binary system at phase 0.25 where the two components would have their 
maximum separation.
\label{fig2}}
\end{figure}

%%Figure 3 -- Light Curves for Encased and Contact Binaries

\begin{figure}
\begin{center}
\epsscale{0.80}
\plotone{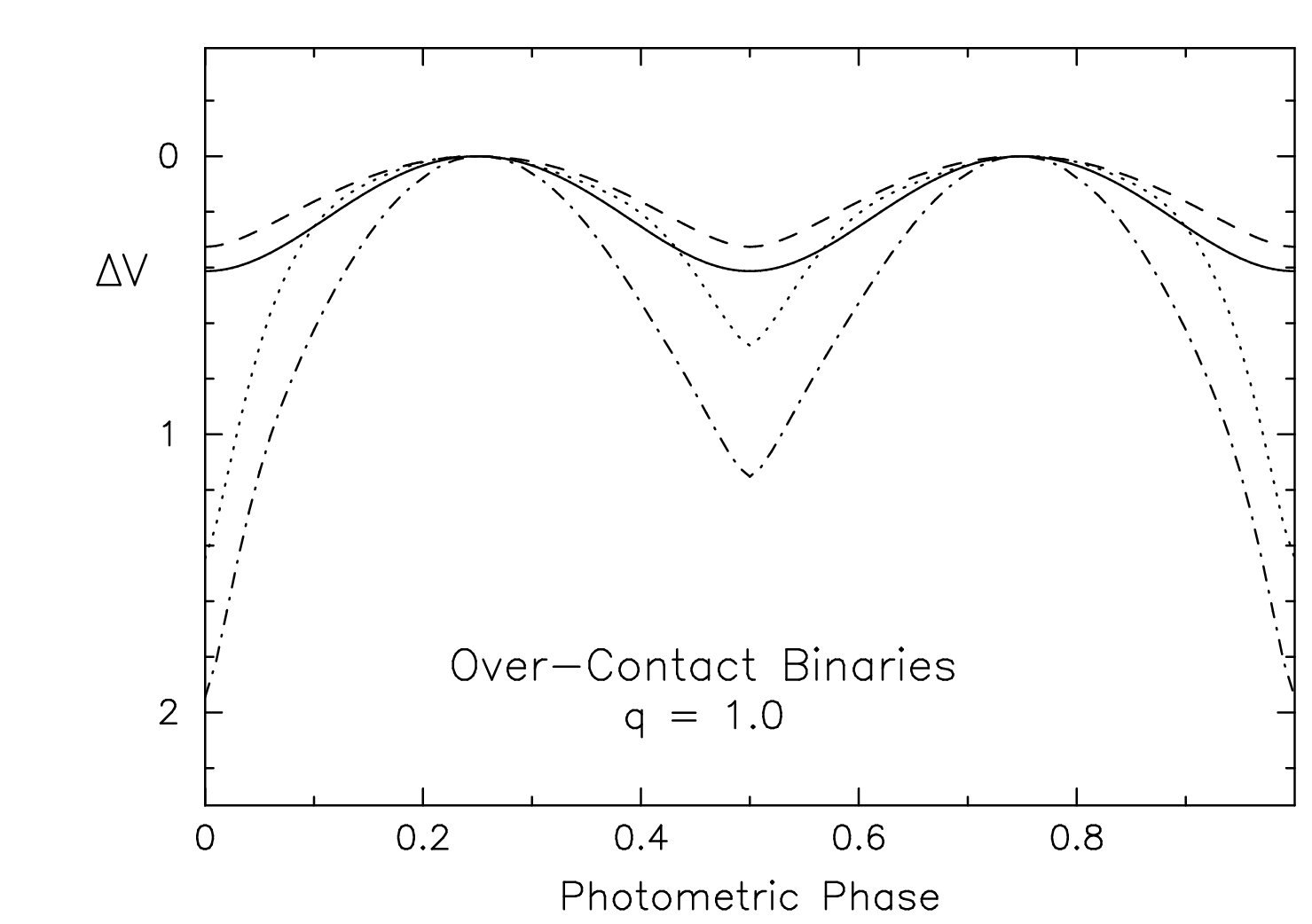}
\end{center}
\caption{Light curves for an encased binary and two limiting contact binaries.
The curves are as follows: solid, non-rotating encased binary with $r_{\rm max}$ = 
1.3$a$ and the expected limb and gravity darkening; dashed, encased binary without 
any limb or gravity darkening; dotted, contact binary with both stars just 
overfilling their Roche lobes (first Lagrangian surface); dot-dashed, contact binary 
filling its outer contact (second Lagrangian) surface.  All these calculations assume 
$i$=90\degr.  
\label{fig3}}
\end{figure}

%%Figure 4 -- Velocity Curve for Encased Binaries

\begin{figure}
\begin{center}
\epsscale{0.80}
\plotone{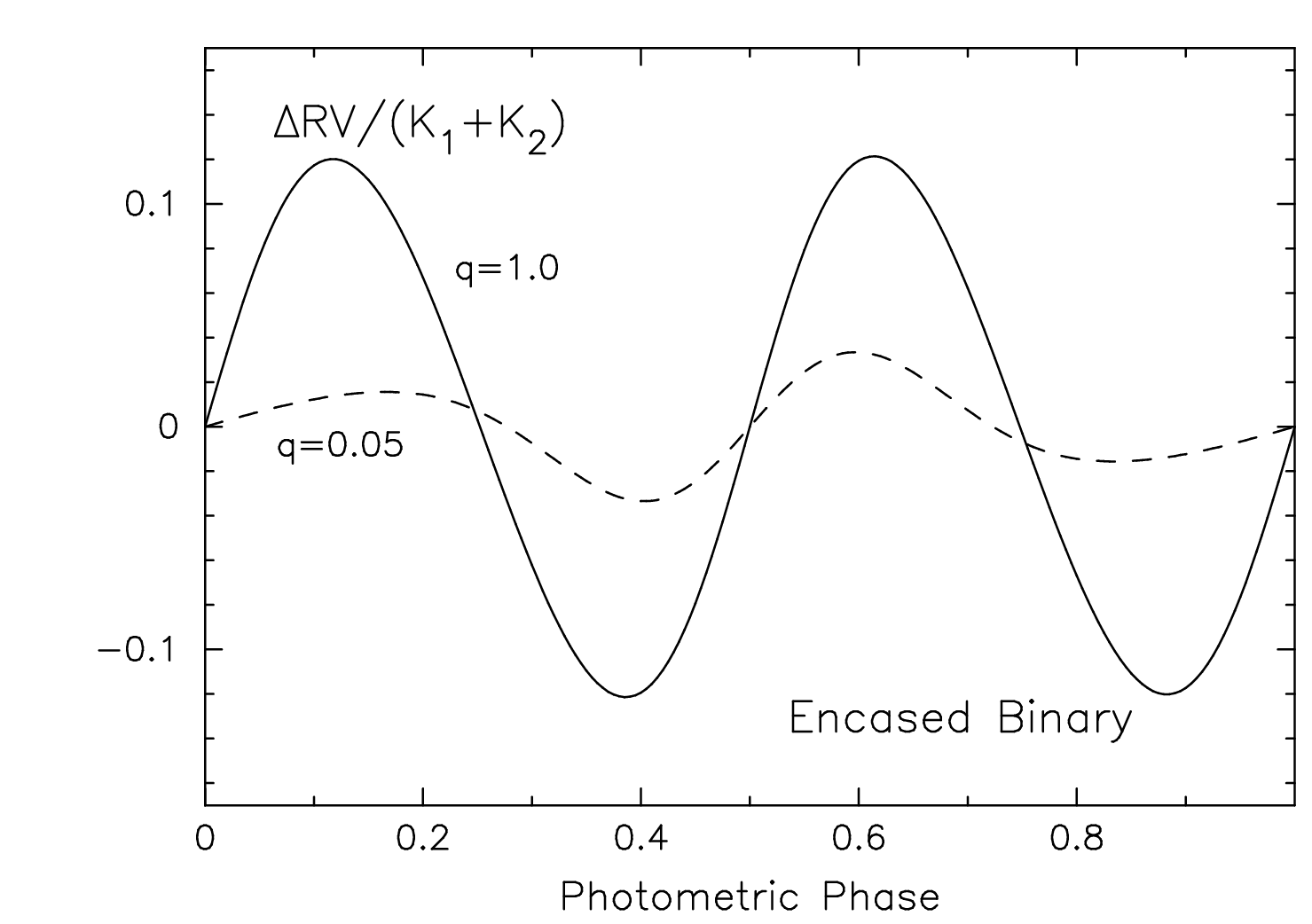}
\end{center}
\caption{Velocity curves calculated for two encased binaries with $R$ = 1.3$a$.
\label{fig4}}
\end{figure}

%%Figure 5 -- Surface Plot for Binary

\begin{figure}
\begin{center}
\epsscale{0.80}
\plotone{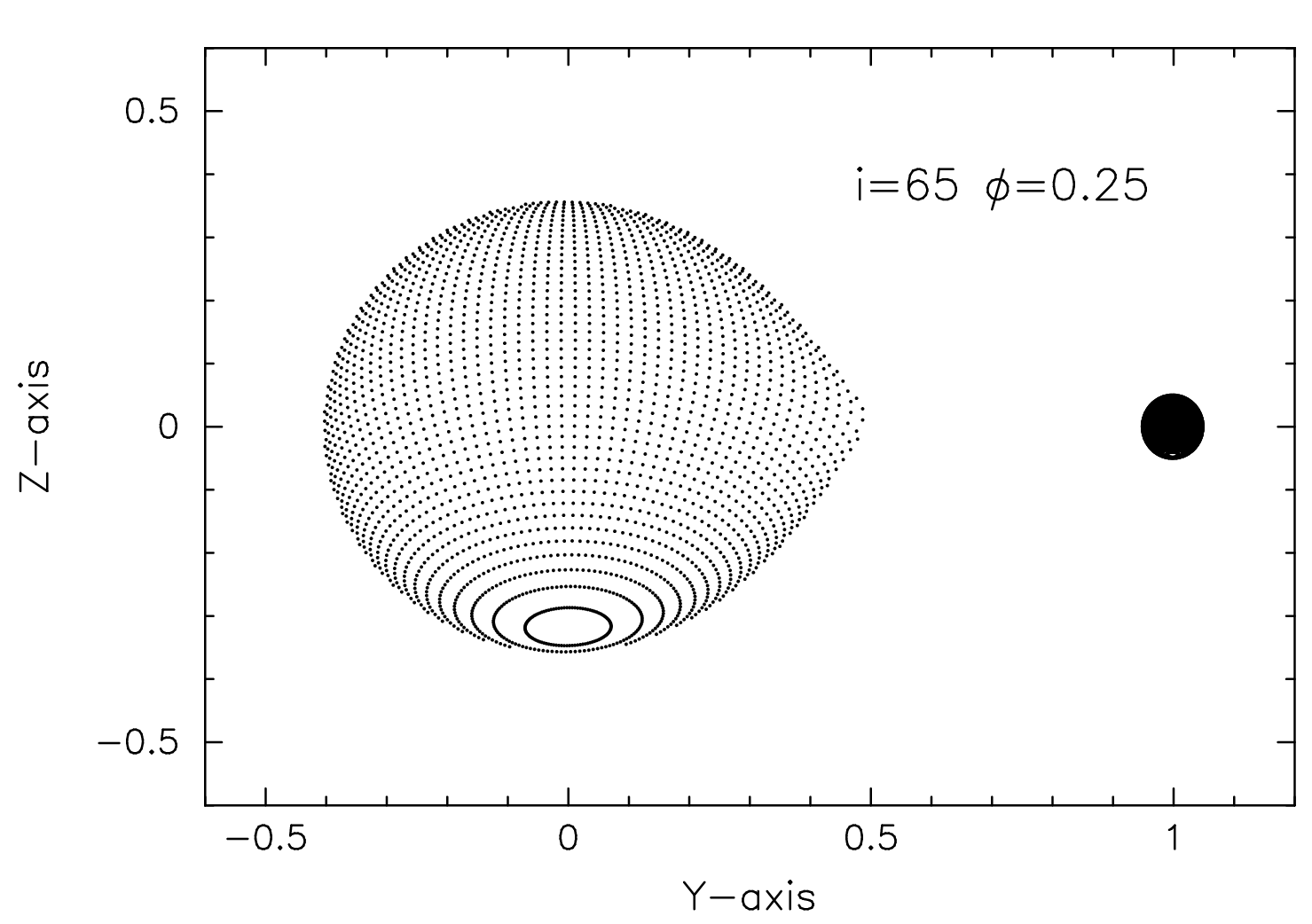}
\end{center}
\caption{The distribution of surface elements in a typical calculation for a 
semi-detached system at phase 0.25 when the bigger, distorted star is approaching.  
The two stars obviously have equal masses in this calculation.
\label{fig5}}
\end{figure}

%%Figure 6 -- Calculated Distortions of RV vs mass ratio

\begin{figure}
\begin{center}
\epsscale{0.80}
\plotone{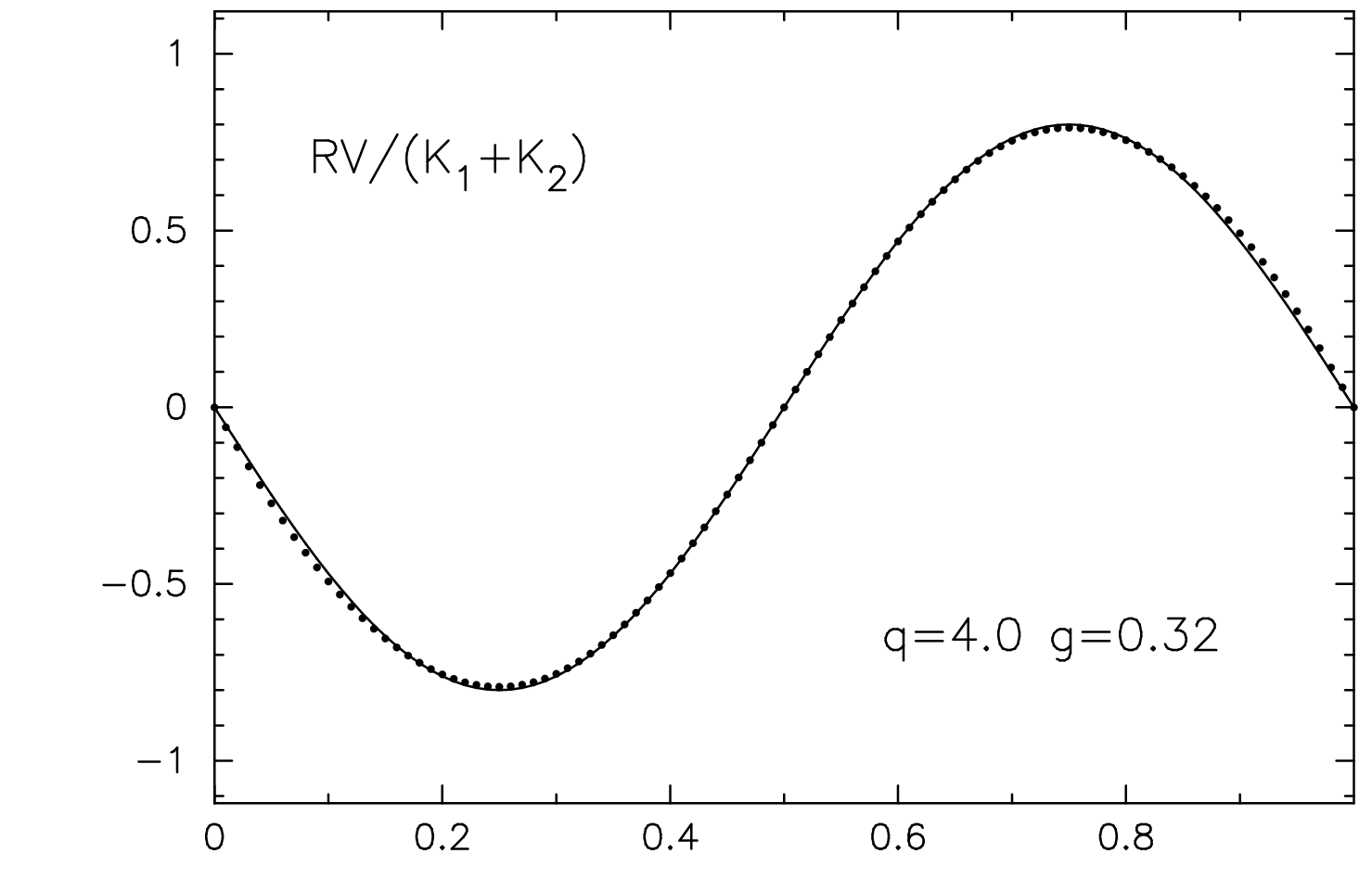}
\plotone{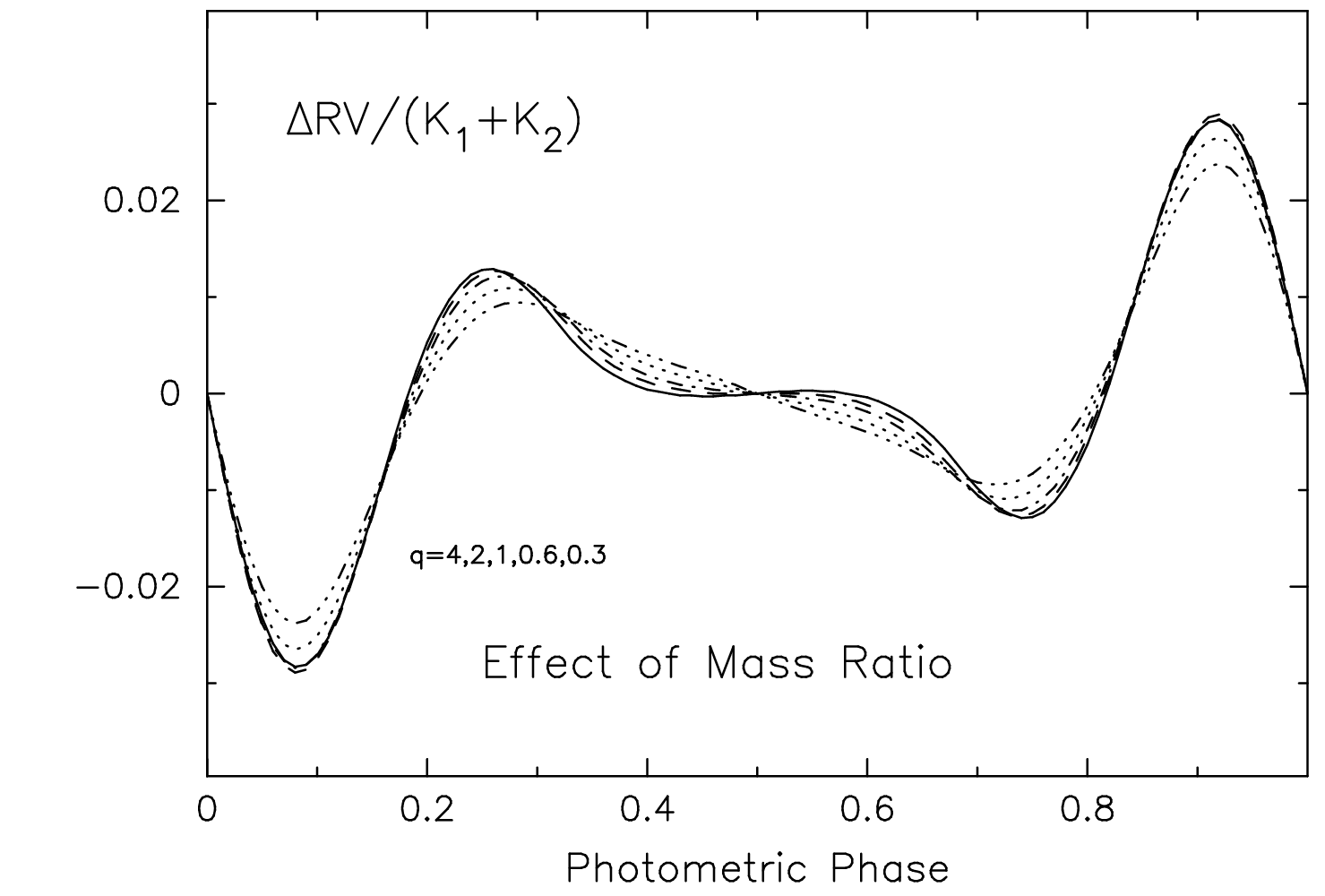}
\end{center}
\caption{Effect of the distortion on a typical velocity curve (at top) 
and some calculated deviations in the lower panel.  The top panel shows a 
calculation for $i$ = 85\degr\ and $q$ = \M2/\M1\ = 1.0, where Star 2 is the 
unseen companion to the visible Star 1.  The dots are calculated velocities 
for the tidally distorted contact component, and the solid curve is the 
sinusoidal velocity of that star's center of mass.  In the bottom panel, 
the five curves give calculations for a range of mass ratio with $i$ = 
85\degr.  The smaller $q$s give roughly the same amplitude, while the larger 
$q$s typical of classical Algol binaries ($q$ = 2.0 and 4.0) give noticeably 
smaller amplitudes.
\label{fig6}}
\end{figure}

%%Figure 7 -- Calculated Distortions of RV vs. Inclination

\begin{figure}
\begin{center}
\epsscale{0.80}
\plotone{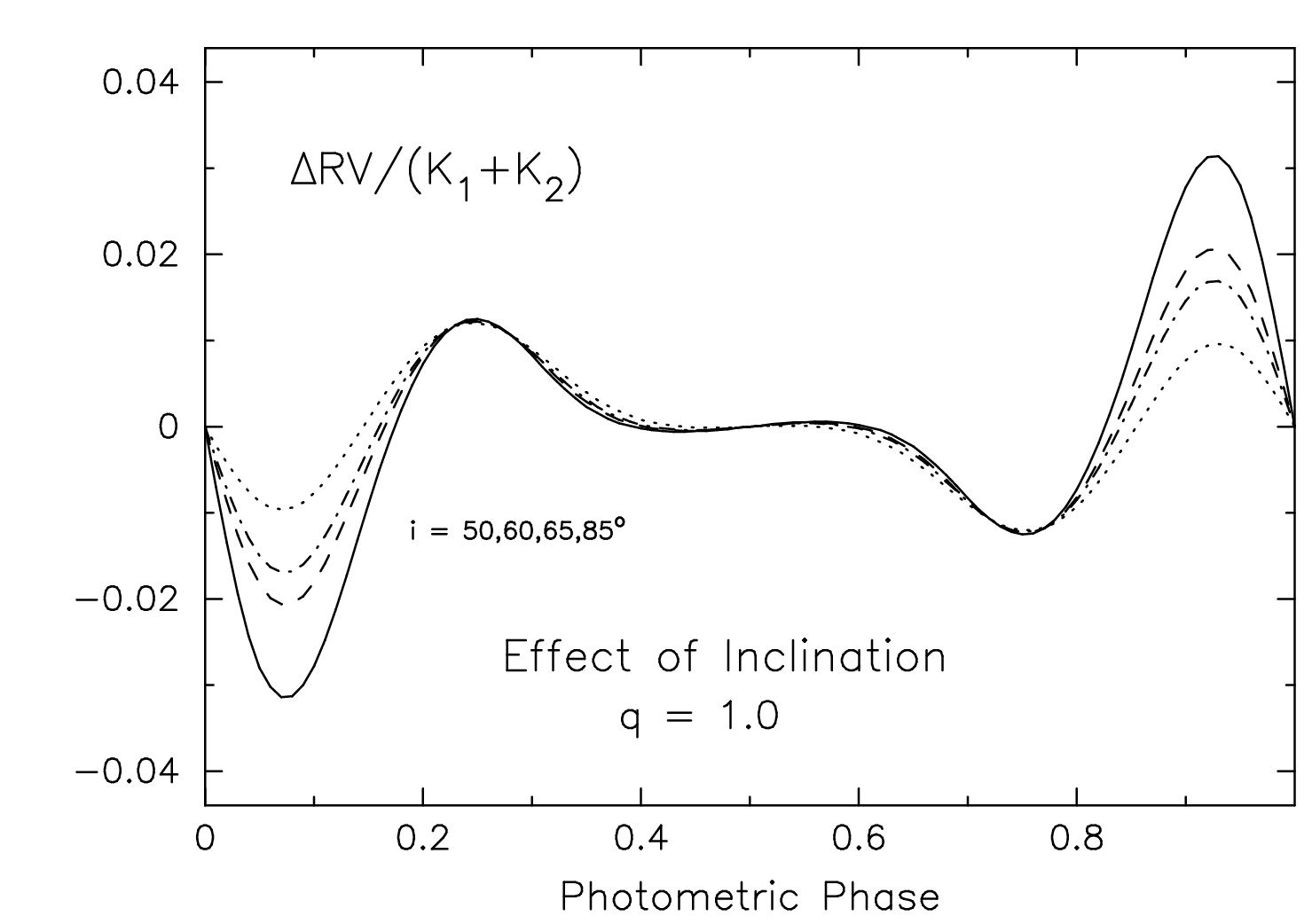}
\end{center}
\caption{Effect of orbital inclination on the distortion of the velocity curve.
Inclination ranges from 50\degr\ (dotted curve) to 85\degr\ (solid curve). Note 
how different this dependence on inclination is for the two conjunctions.
\label{fig7}}
\end{figure}

%%Figure 8 -- Calculated Distortions of RV vs. Gravity Darkening

\begin{figure}
\begin{center}
\epsscale{0.80}
\plotone{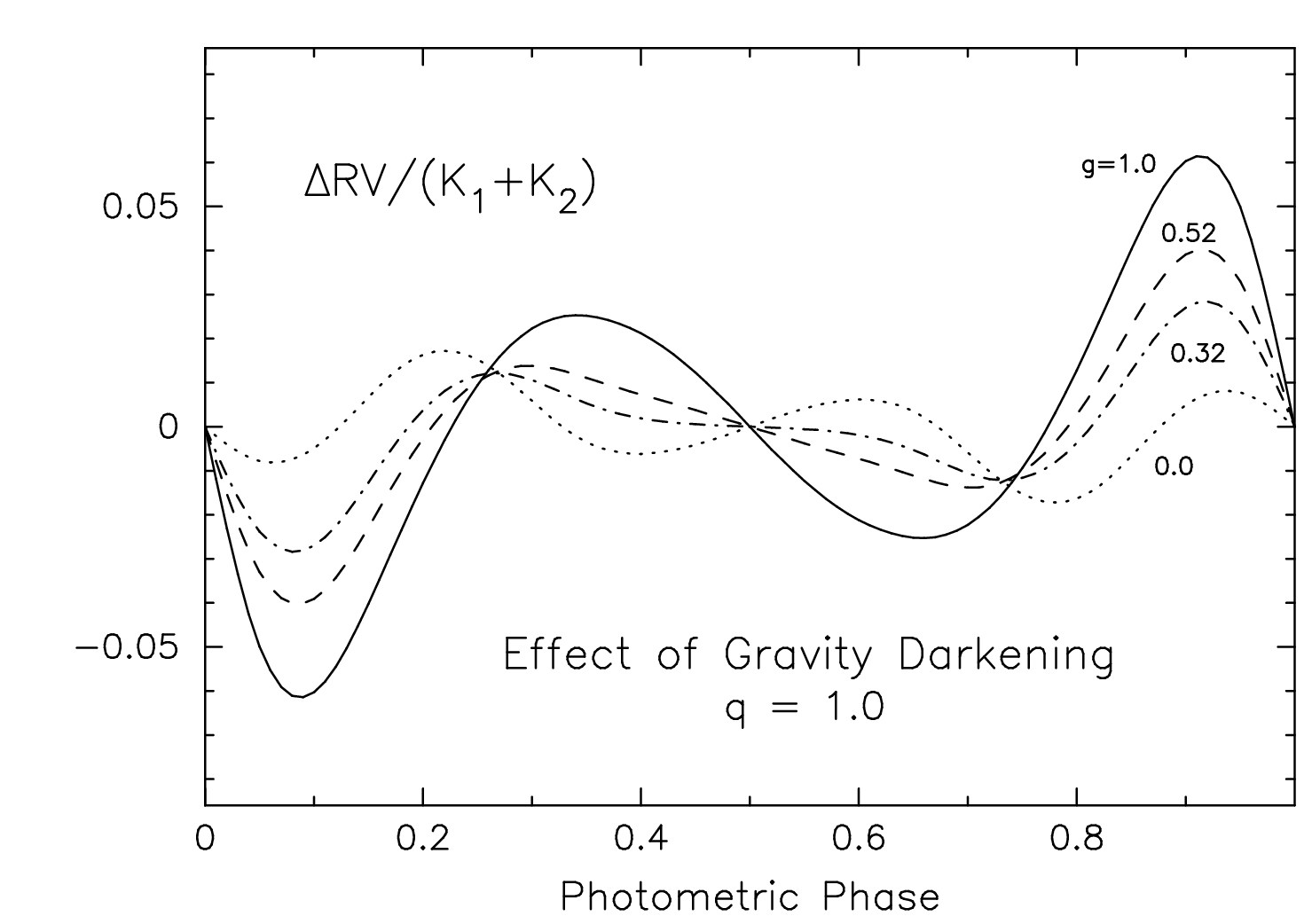}
\end{center}
\caption{Effect of gravity darkening on the distortion of the velocity curve.
Curves are as follows: dotted for $g$=0.0 (no gravity darkening), dot-dashed for 
$g$=0.32 (Lucy's theoretical value), dashed for $g$=0.52, and solid for $g$=1.0
(radiative--von Zeipel--gravity darkening).
\label{fig8}}
\end{figure}

%%Figure 9 -- Orbital Solution for 5 Ceti

\begin{figure}
\begin{center}
\epsscale{0.80}
\plotone{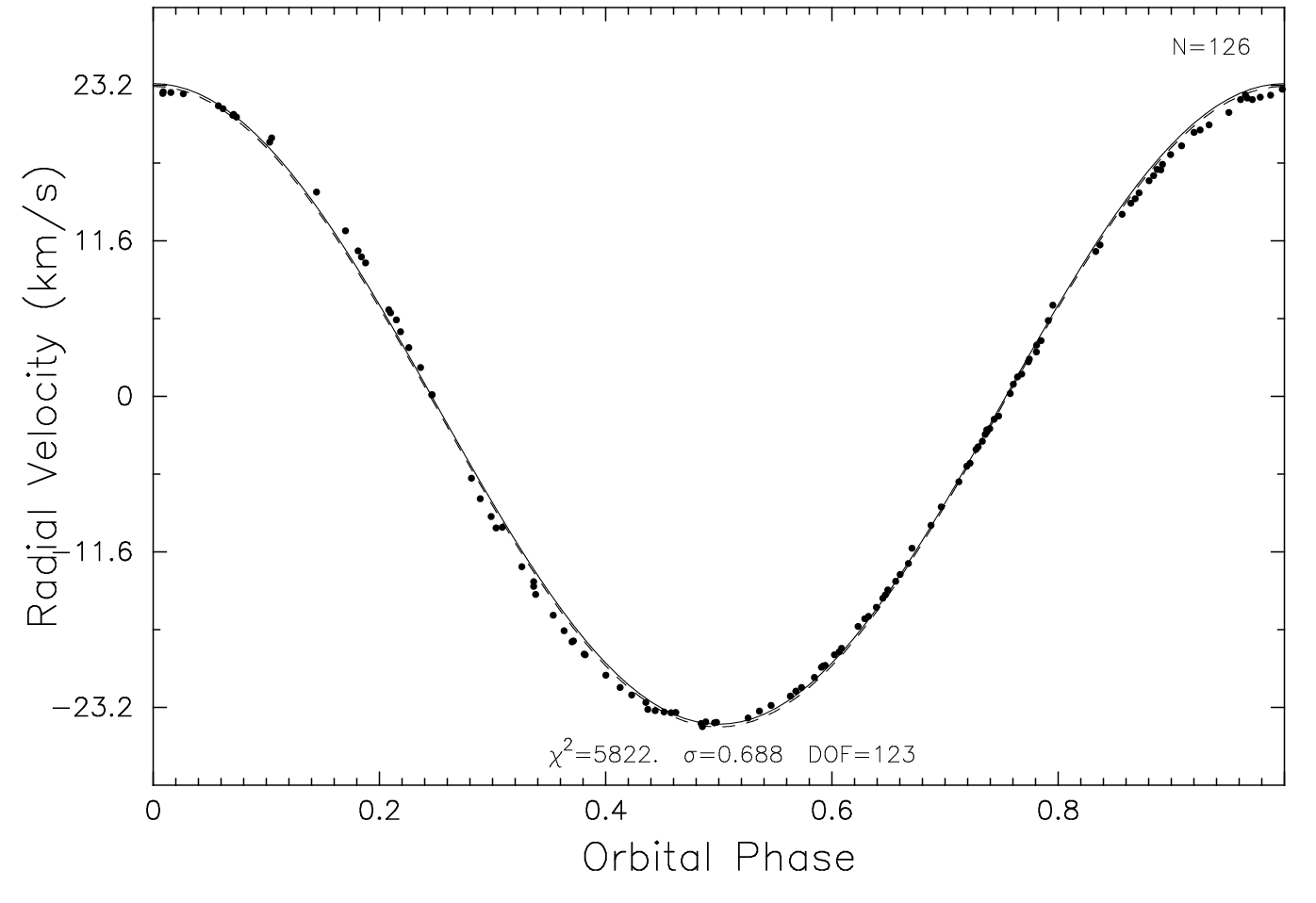}
\end{center}
\caption{Fit of a circular orbit to measured velocities for 5~Cet.  The 
RMS deviation for this fit is $\sigma$ = 0.69 \kmps; the systematic deviation 
is obvious.  Compare this graph to the upper panel of Figure 6.
\label{fig9}}
\end{figure}

%%Figure 10 -- Deviations for 5 Ceti 

\begin{figure}
\begin{center}
\epsscale{0.80}
\plotone{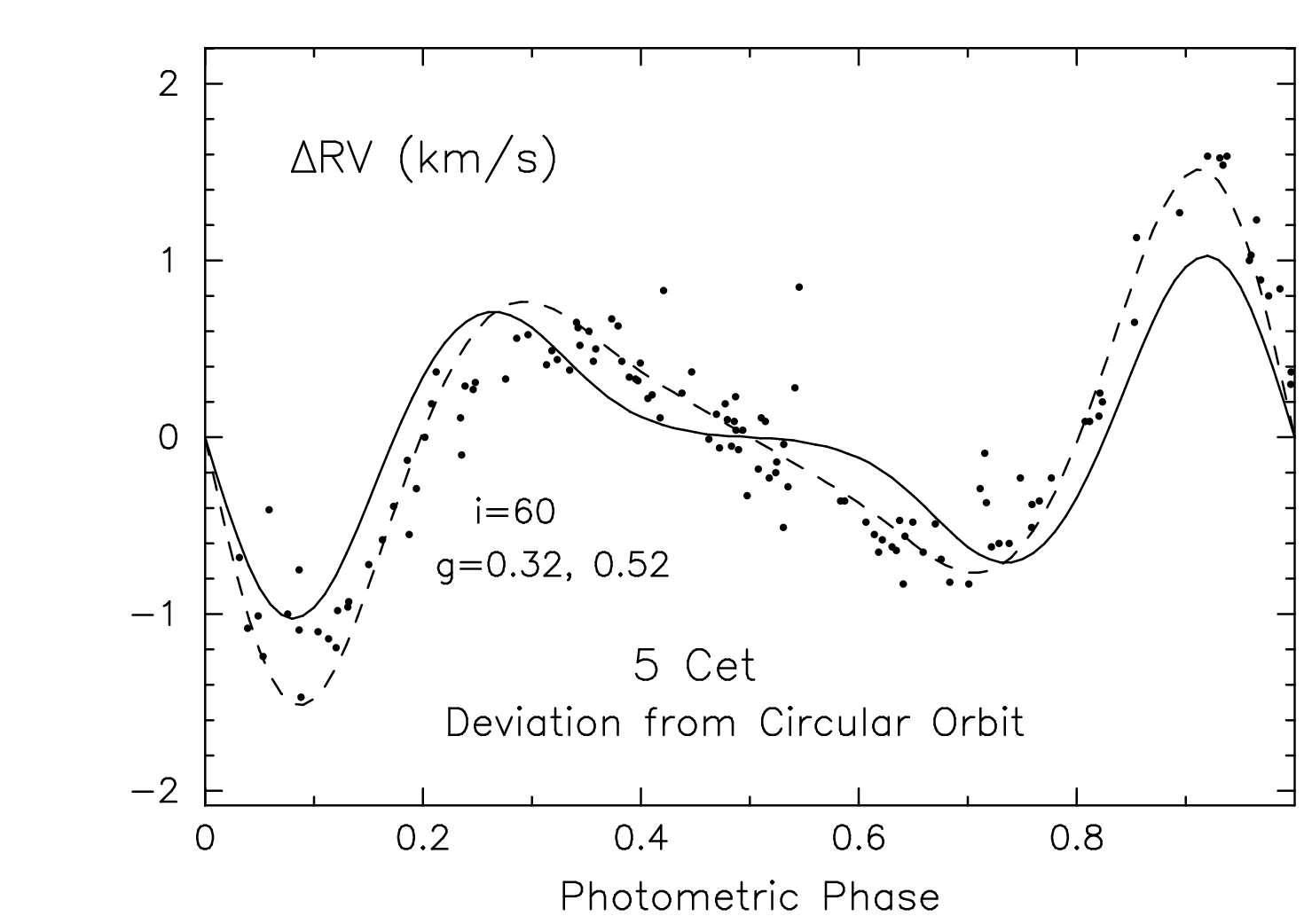}
\end{center}
\caption{Deviation of measured radial velocities of 5~Cet from a sine curve.
Compare the shape of this curve to the lower panel in Figure 6.  Dots are the 
measured deviations, the differences between the observations and fitted curve 
in Figure 9.  Curves give theoretically calculated deviations for two values 
of the gravity exponent, solid for $g$=0.32, dashed for $g$=0.52.  Both 
calculations assume the mass ratio, 0.7, derived from line broadening.
\label{fig10}}
\end{figure}

%%Figure 11 -- Deviations for AX Mon 

\begin{figure}
\begin{center}
\epsscale{0.80}
\plotone{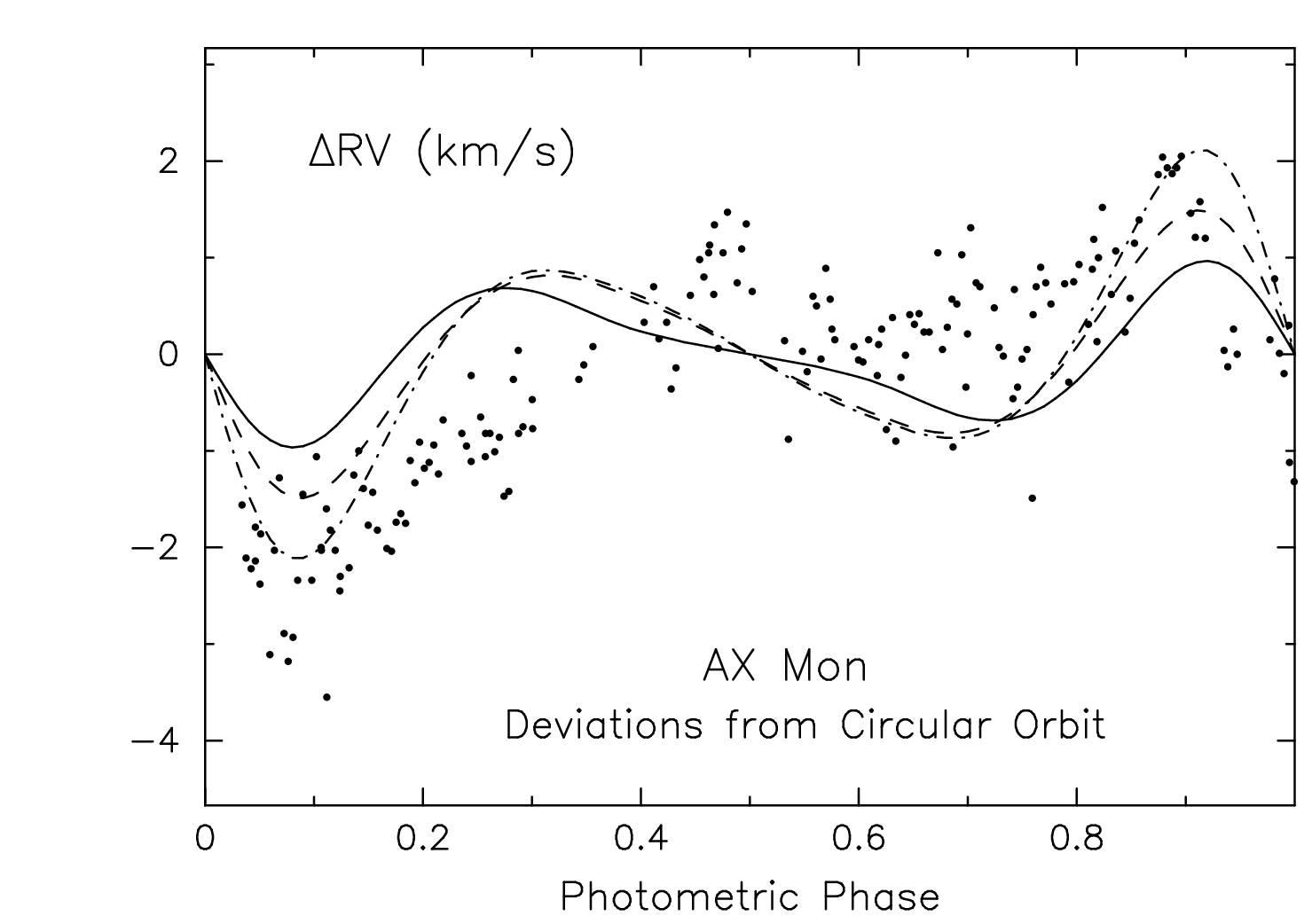}
\end{center}
\caption{Deviation of measured radial velocities of AX~Mon from a sine curve.
Compare the shape of this curve to Figure 6.  As in Figure 10, dots are the 
differences between observed velocities and a fitted sine curve.  The curves give 
theoretically expected variations for $g$=0.32 with $i$=60\degr\ (solid), \
$g$=0.52 with $i$=60\degr\ (dashed), $g$=0.52 with $i$=75\degr\ (dot-dashed).
All assume $q$=3.0 and $x$=0.77.
\label{fig11}}
\end{figure}

%%Figure 12 -- Deviations for TT Hya 

\begin{figure}
\begin{center}
\epsscale{0.80}
\plotone{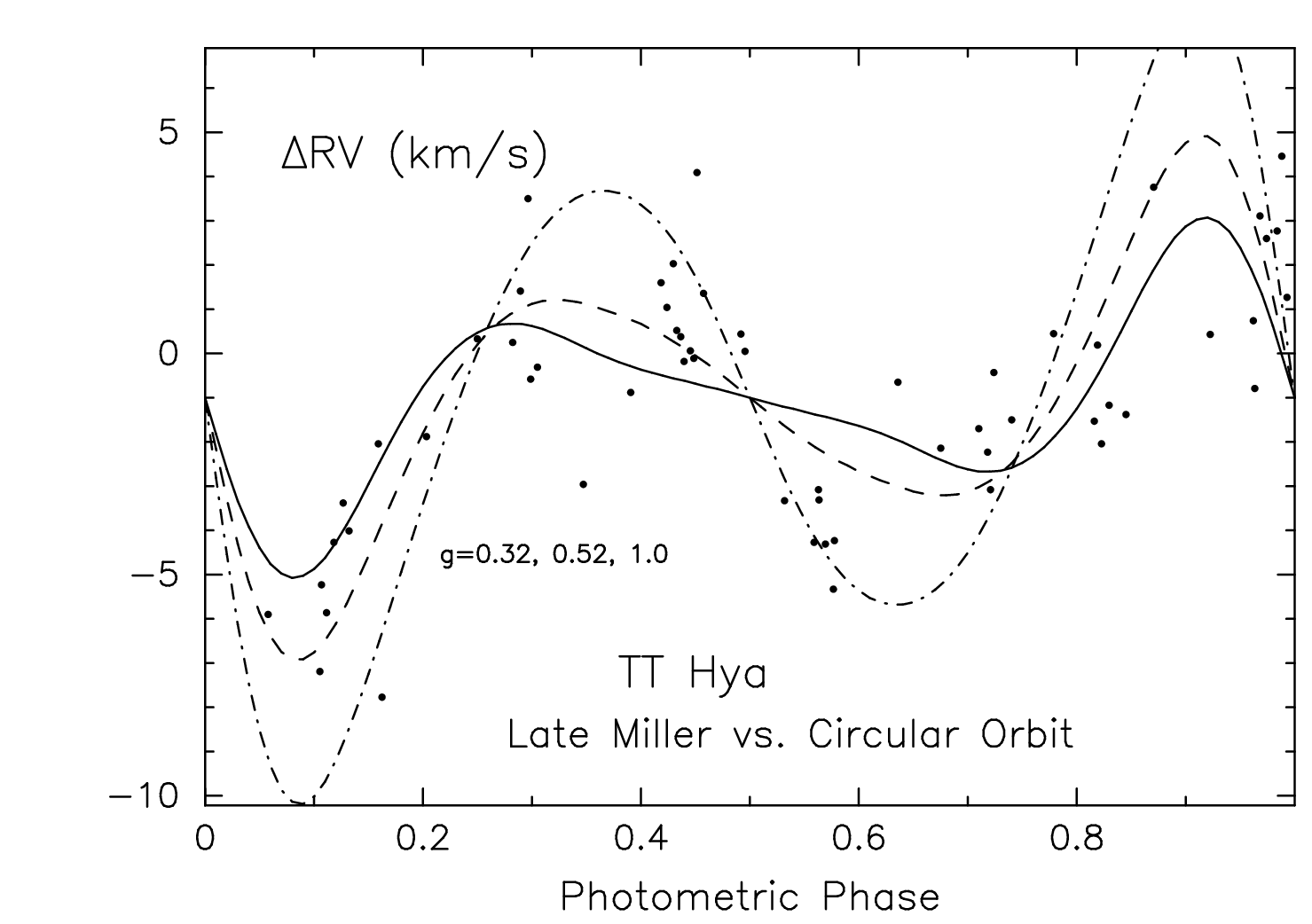}
\end{center}
\caption{Deviation of measured radial velocities of TT~Hya (the selected data from 
Miller et al.\ [2007] identified in \S\ 3.2) from a sine curve.  As in Figure 10, 
dots are the differences between observed velocities and a fitted sine curve.  The 
curves give theoretically expected variations for $g$=0.32 (solid), $g$=0.52 (dashed), 
and $g$=1.0 (dot-dashed).
\label{fig12}}
\end{figure}

\end{document}